\documentclass[aps,prl,twocolumn,superscriptaddress,showpacs]{revtex4}

\usepackage{graphicx}

\begin{document}

\title{Scaling of the anomalous Hall effect in Sr$_{1-x}$Ca$_x$RuO$_3$}

\author{R. Mathieu\cite{email,cryo}}

\affiliation{Spin Superstructure Project (ERATO-SSS), JST, AIST Central 4, Tsukuba 305-8562, Japan}

\author{A. Asamitsu\cite{cryo}}
\affiliation{Spin Superstructure Project (ERATO-SSS), JST, AIST Central 4, Tsukuba 305-8562, Japan}
\affiliation{Department of Applied Physics, University of Tokyo, Tokyo
113-8656, Japan}

\author{H. Yamada}
\affiliation{Correlated Electron Research Center (CERC), AIST Central 4, Tsukuba 305-8562,
Japan}

\author{K. S. Takahashi}
\affiliation{Department of Applied Physics, University of Tokyo,
Tokyo 113-8656, Japan}

\author{M. Kawasaki}
\affiliation{Correlated Electron Research Center (CERC), AIST Central 4, Tsukuba 305-8562,
Japan}
\affiliation{Institute for Materials Research, Tohoku University,
Sendai 980-8577, Japan}

\author{Z. Fang}
\affiliation{Spin Superstructure Project (ERATO-SSS), JST, AIST Central 4, Tsukuba 305-8562, Japan}
\affiliation{Institute of Physics, Chinese Academy of Science, Beijing 100080, China}

\author{N. Nagaosa}
\affiliation{Department of Applied Physics, University of Tokyo,
Tokyo 113-8656, Japan}
\affiliation{Correlated Electron Research Center (CERC), AIST Central 4, Tsukuba 305-8562,
Japan}
\affiliation{CREST, Japan Science and Technology Agency (JST)}

\author{Y. Tokura}
\affiliation{Spin Superstructure Project (ERATO-SSS), JST, AIST  Central 4, Tsukuba 305-8562, Japan}
\affiliation{Department of Applied Physics, University of Tokyo,
Tokyo 113-8656, Japan}
\affiliation{Correlated Electron Research Center (CERC), AIST Central 4, Tsukuba 305-8562,
Japan}

\date{\today}

\begin{abstract}
The anomalous Hall effect (AHE) of ferromagnetic thin films of Sr$_{1-x}$Ca$_{x}$RuO$_3$ (0 $\leq x \leq$ 0.4)
is studied as a function of $x$ and temperature $T$.
As $x$ increases, both the transition temperature $T_c$ and the 
magnetization $M$ are reduced and vanish near $x \sim$ 0.7. 
For all compositions, the transverse resistivity $\rho_{H}$
varies non-monotonously with $T$, and even changes sign, thus
violating the conventional expression 
$\rho_{H}=R_o B + 4\pi R_s M(T)$ ($B$ is the magnetic induction,  while $R_o$ and $R_s$ are the ordinary and anomalous Hall coefficients).
From the rather complicated data of $\rho_H$, 
we find a scaling behavior of the transverse conductivity 
$\sigma_{xy}$ with $M(T)$, which is well reproduced by the first-principles band calculation assuming the intrinsic origin of the AHE.
\end{abstract}

\pacs{75.30.-m, 72.15.-v}
\maketitle
It has been known that the Hall resistivity $\rho_H$ in a
ferromagnet \cite{Hall} has some extra contribution originated 
from the spontaneous magnetization, which is assumed and often
observed experimentally to be fitted by 
$\rho_{H}=R_0 B + 4\pi R_sM$, where $B$ is the magnetic induction and
$M$ is the magnetization of the material. $R_0 B$ represents the ordinary 
Hall contribution, which is related to the nature and amount of charge 
carriers. It is a linear function of the applied magnetic field $H$ as 
in the Hall measurement geometry, $B$=$H$. $R_s M$ is referred to as 
the anomalous Hall term, and is usually associated with the spin
polarization of the conduction carriers and the 
relativistic spin-orbit interaction. 
According to the above definition of $\rho_{H}$, the anomalous Hall 
term is proportional to the magnetization of the material.
However the quantitative analysis of the AHE has seldom been
completed since its mechanism has not yet been established, and 
theories give much smaller values compared with the
experiments. Furthermore, most of the theories regard the AHE
as from extrinsic origins, involving processes such as skew scatterings 
\cite{skew} and side-jump mechanisms\cite{side}. 
Therefore the magnitude of the AHE
depends on the concentration and scattering strength of impurities, 
thermal spin-agitation, etc.
In contrast to these extrinsic mechanisms, several 
works\cite{kl,ye,nagaosa,salamon,jungwirth} regard
 AHE of the intrinsic origin.  Namely
the phase of the Bloch wave-function in the momentum space 
determines the Hall conductivity $\sigma_{xy}$, which is 
largely determined by the band crossing points acting as 
``magnetic monopoles''\cite{Fang}. As an explicit example,
the AHE and magneto-optical effect of SrRuO$_3$ have been studied, 
and a good agreement was obtained between 
theory and experiments\cite{Fang}. In the present paper, we report the extensive study of the AHE in  Sr$_{1-x}$Ca$_{x}$RuO$_3$ (0 $\leq x \leq$ 0.4) as a function of $T$ and $x$ to reveal its systematics. We have found the 
scaling behavior of the transverse conductivity $\sigma_{xy}$ in terms of the $T$- and $x$-dependence of the magnetization $M(x,T)$,
which is in fairly good agreement with the 
first-principles band calculation. 
This gives a firm evidence for the intrinsic origin of the AHE.

Several members of the Ruddelson-Popper-type Sr$_{n+1}$Ru$_{n}$O$_{3n+1}$ series show metallic properties, as well as superconductivity and magnetic order\cite{SROxyz}. SrRuO$_3$ ($n=\infty$, perovskite) is ferromagnetic\cite{SRO1,SRO2} with  a Curie temperature ($T_c$) around 160 K, and a fairly large spin orbit coupling. The $4d$ orbitals of Ru$^{4+}$ are rather extended, and the Coulomb repulsion is small compared to the 
band width. The ferromagnetic properties of SRO are usually associated 
with a narrow itinerant band resulting from the hybridization between the 
Ru($t_{2g}$) and O($2p$) orbitals. 
  
While similar structurally, and as well metallic, CaRuO$_3$ (CRO) does not exhibit ferromagnetism and its magnetic state is still under discussion\cite{CRO}. In 
Sr$_{1-x}$Ca$_x$RuO$_3$, the ferromagnetic interaction 
becomes weaker with increasing $x$\cite{SCRO}. For the compounds with larger 
Ca-doping ($x$ $\geq$ 0.7), no clear phase transition is discerned, 
and only some irreversibility is observed in the magnetization curves of 
these materials.
The disappearance of the long range magnetic order is commonly related to 
the distortion of the RuO$_6$ octahedra associated with the partial or total replacement 
of Sr by Ca, and the corresponding narrowing of the $4d$ bandwidth\cite{SCRO2}.

It is difficult to grow clean single crystals of Ca-doped SRO, 
although it is possible to prepare high quality single crystals of the 
end compounds, SRO and CRO. On the other hand, it is nowadays possible to grow high quality 
epitaxial films of SrRuO$_3$ and Sr$_{1-x}$Ca$_{x}$RuO$_3$\cite{eom}. In the present article, we study the anomalous Hall 
resistivity of epitaxial films of 
Sr$_{1-x}$Ca$_{x}$RuO$_3$ (0 $\leq x \leq$ 0.4), and its 
evolution as the ferromagnetic interaction decreases with $x$. 
The Hall resistivity contains an anomalous component, associated with the 
ferromagnetic ordering of the samples at low temperatures. This component is 
not simply proportional to the $M$, as usually 
assumed. The results reveal a close relation between 
$\sigma_{xy}$  and the spin polarization of the system. Such a 
correlation was predicted by first-principles calculations taking account of the spin-orbit interaction in terms of the Berry phase connection\cite{Fang}. The calculations successfully reproduce the non-monotonous variation of $\sigma_{xy}$ 
with temperature (via its magnetization), as well as its sign change. 
The effects of disorder and structural changes on the anomalous 
conductivity are discussed.

\begin{figure}[h]
\includegraphics[width=0.46\textwidth]{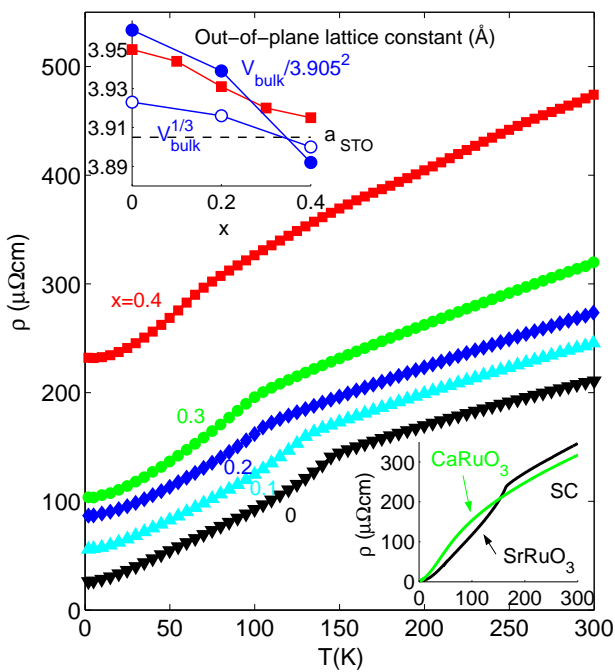}
\caption{(color online) Temperature dependence of the resistivity of the Sr$_{1-x}$Ca$_{x}$RuO$_3$ films (main frame) and single crystals of SrRuO$_3$ and CaRuO$_3$ (lower inset). The upper 
inset shows the variation of the out-of-plane lattice parameter of the 
films (squares). The lattice parameter of the SrTiO$_3$ substrate $a_{STO}$ is 
indicated, as well as the average lattice parameter obtained for  Sr$_{1-x}$Ca$_{x}$RuO$_3$ polycrystalline samples, for $a$=$b$=$c$ ($V^{1/3}$, $V$ is the unit cell volume), and for a perfect elastic strain (Poisson ratio of 0.5; $V/a_{\rm STO}^2$). 
The parameters of $x$ = 0, 0.1, and 0.2 ideally lie between the $V^{1/3}$ 
and $V/a_{\rm STO}^2$ values; the small deviation observed from $x$ = 0.3 
and 0.4 may be related to a minor Ru deficiency; ruthenium oxides are very 
robust against oxygen deficiencies, so that the oxygen content should be stoichiometric.}
\label{figRT}
\end{figure}
\indent Thin ($\sim$ 500 {\AA}) films of Sr$_{1-x}$Ca$_{x}$RuO$_3$ 
($x$ = 0, 0.1, 0.2, 0.3, and 0.4) were epitaxially grown on the (001) 
surfaces of high quality SrTiO$_3$ single-crystal substrates\cite{kawasaki} by pulsed laser deposition 
(PLD). Bulk single crystal SrRuO$_3$ and CaRuO$_3$ were prepared 
using a flux method for comparison. The quality of the films and 
phase-purity of 
the single crystals were confirmed by x-ray diffraction. Both single 
crystals have orthorhombic structure;  SrRuO$_3$ has a relatively small 
orthorhombicity ($c/a$=1.003, $c/b$=0.996), while it is larger for 
CaRuO$_3$ ($c/a$=1.01, $c/b$=0.980). The epitaxial thin films are coherently strained 
by the SrTiO$_3$ substrate, yielding a tetragonal distortion in the [001] 
direction ($c/a$=$c/b$=1.01 for $x$=0). As a result, the out-of-plane 
lattice constants of the films are elongated (c.f. inset of Fig.~\ref{figRT}), 
and due to the spin-orbit interaction, the easy axis of magnetization is 
perpendicular to the film plane\cite{strain2}. Magnetic and transport measurements were 
performed on the Sr$_{1-x}$Ca$_{x}$RuO$_3$ thin films and the single 
crystals of SRO and CRO. The magnetization data was recorded on a MPMS5S SQUID magnetometer 
using a magnetic field applied normal to the plane of the films. 
The films were then patterned in a six-lead Hall bar geometry using 
conventional photo-lithography and Ar ion etching for transport measurements. 
The Hall resistivity  $\rho_{H}$ was measured with a PPMS6000 system together 
with the longitudinal resistivity $\rho_{xx}=\rho$ as a function of $H$ and $T$. The anomalous  resistivity  
$\rho_{xy}$  was extrapolated to $H$ = 0 from $\rho_{H}$  vs $H$ 
measurements up to $H$ = $\pm$ 9 T at constant temperatures (from 2 K to 200 K) after subtraction of the ordinary Hall contribution,  and  the transverse conductivity 
$\sigma_{xy}$ was estimated as -$\rho_{xy}/\rho^2_{xx}$. A small (as the patterned leads are nearly symmetric) magnetoresistance was removed by subtracting $\rho_{H}$($-H$) to $\rho_{H}$($H$). 
First-principles calculations of $\sigma_{xy}$ were performed 
assuming orthorhombic and cubic crystal structures. The plane-wave 
pseudo-potential calculations were performed based on the local spin density 
approximation (LSDA), and the spin-orbit coupling was treated 
self-consistently by using the relativistic fully separable 
pseudo-potentials in the framework of non-collinear magnetism formalism. 
The finite life-time broadening  was estimated from the experimental 
residual resistivity and the extended Drude analysis of the longitudinal 
conductivity\cite{Fang}.

\begin{figure}[h]
\includegraphics[width=0.46\textwidth]{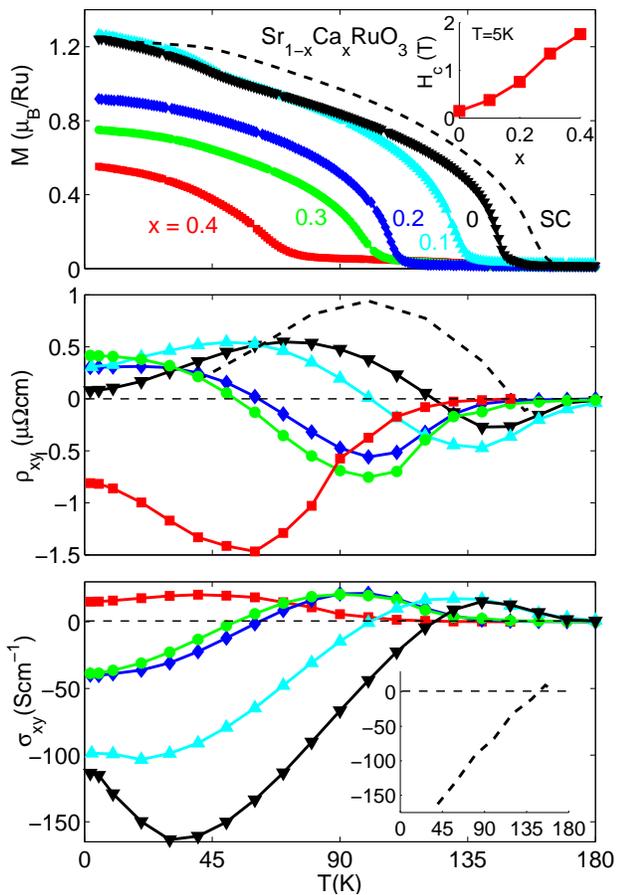}
\caption{(color online) Temperature dependence of the magnetization (top panel), transverse 
resistivity (middle panel), and transverse conductivity (bottom panel) of the Sr$_{1-x}$Ca$_{x}$RuO$_3$ films. In the SQUID experiments, the samples are cooled in $H$= 7 T, 
and the magnetization is recorded on re-heating in $H$ = 0.05 T applied 
normal to the film planes. The results for the single crystal (SC) 
are included for comparison in dashed lines. The top inset shows the monotonous increase 
of the coercivity $H_c$ with increasing Ca doping; $H_c$  was determined from $M-H$ measurements at $T$ = 5 K.}
\label{figMRXYT}
\end{figure}
\indent All the investigated samples are metallic as seen in Fig.~\ref{figRT} and inset. 
The residual resistivity at low temperatures is very low 
($\sim$ 2.5 $\mu\Omega$cm) for the single crystals, and increases for the films from 26 ($x$ = 0) to 232  $\mu\Omega$cm ($x$ = 0.4). A kink is observed in the resistivity curves, around the paramagnetic-to-ferromagnetic transition temperatures ($T_c$) of the films. No long range magnetic order is observed in CRO, and as seen in inset, the resistivity curve has no anomaly in the measured range of temperature. Due to the above mentioned strain effects, the epitaxial film of SRO has a slightly lower $T_c$ than the single crystal, near 150 K\cite{strain2,strain}. The substitution of Sr by Ca in Sr$_{1-x}$Ca$_{x}$RuO$_3$ weakens the ferromagnetic interaction, and $T_c$ is greatly reduced. It is reduced to $\sim$ 110 K for $x$ = 0.2, and $\sim$ 70 K for $x$ = 0.4.

The top panel of Fig.~\ref{figMRXYT} shows the temperature dependence of the 7 T-cooled magnetization of the films probed in a small field of $H$ = 0.05 T applied normal to the plane, i.e. in the same direction as in the Hall measurements, along the easy axis of magnetization. As seen in inset, the coercivity $H_c$ of the films monotonously increases with increasing Ca doping. In addition to possible disorder, the Ca substitution yields structural changes, affecting the magnetic properties and, as we will discuss below, the AHE.  At low temperatures,  $M$($H$ = 7 T) amounts to $\sim$ 1.5  $\mu_B$/Ru for the undoped SRO film. Due to the itinerant character of the magnetism, the obtained moment is smaller than expected according to Hund's rule (2 $\mu_B$/Ru; $S$=1). $M$($H$ = 7T) decreases with increasing Ca doping, and amounts to $\sim$ 0.75 $\mu_B$/Ru for $x$ = 0.4. The middle panel of Fig.~\ref{figMRXYT} shows the temperature dependence of the anomalous Hall resistivity $\rho_{xy}$ for all the films and the SRO single crystal. At a constant temperature, $\rho_H$ is a linear function of $H$ at high fields, with a negative proportionality constant ($R_o <$ 0), indicating charge carriers of electron-like nature.
\begin{figure}[h]
\includegraphics[width=0.46\textwidth]{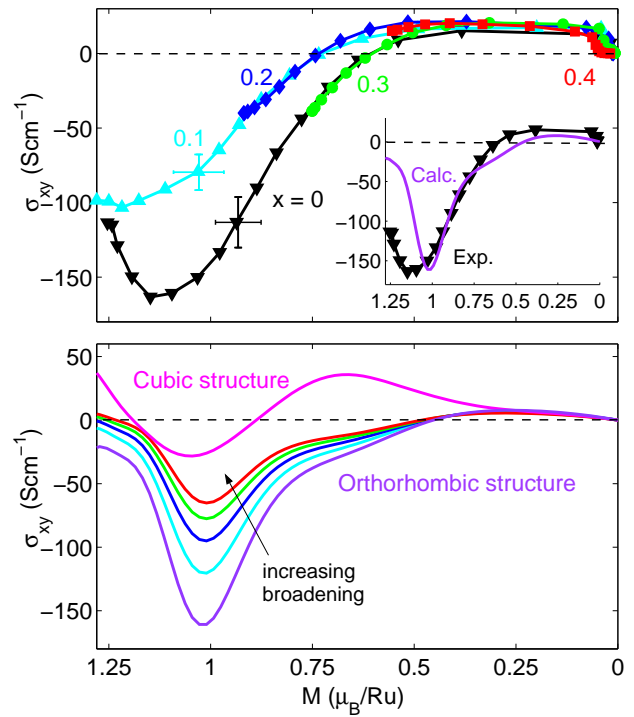}
\caption{(color online) Top panel: The transverse conductivity $\sigma_{xy}$ data obtained for the films is plotted against the magnetization $M$, using the data from Fig.~\ref{figMRXYT}. Typical errors bars are indicated. As the areas of the films are well defined by patterning, the uncertainty on the magnitude of $\rho$ is mainly determined by the error in thickness determination, which amounts to $\sim$  5 \% of the actual thickness. Including measurement and determination errors, there is an uncertainty of $\sim$ 15 \% on $\sigma_{xy}$ . The uncertainty on the magnitude of $M$ is, as $\rho$, of $\sim$ 5 \%. Bottom panel: First -principles calculations for cubic and orthorhombic structures. Results obtained using different broadening parameters are shown in the orthorhombic case.}
\label{figSXYT}
\end{figure}
As seen in the figure, $\rho_{xy}$ of the films varies non-monotonously with $T$. No AHE is observed at high temperatures. The AHE appears just above $T_c$;  $\rho_{xy}$ is negative, and show a maximum near $T_c$. At lower temperatures,  $\rho_{xy}$ changes sign (near 120 K for $x$ = 0, and  60 K for $x$ = 0.2), and remain positive down to the lowest temperature. For $x$ = 0.4 with a lower $T_c$, however, the sign change is not observed. Similar features are observed for the single crystal of SRO; the resistivity of the single crystals was too low below 40 K to estimate $\rho_{xy}$ (and thus $\sigma_{xy}$). As seen in the bottom panel of Fig.~\ref{figMRXYT}, the anomalous conductivity remains fairly large at low temperatures, amounting to $\sim$ -100 Scm$^{-1}$ at 2 K for the undoped and $x$ = 0.1 films.  $\sigma_{xy}$($T$) shows a similar high-temperature peak, which, as $T_c$, is shifted to lower temperatures as the Ca doping increases.  If the anomalous conductivity data in Fig.~\ref{figMRXYT} is plotted against the magnetization instead of the temperature, as in the top of Fig.~\ref{figSXYT}, one observes, within the measurement uncertainties, a similar or universal behavior of the Hall conductivity for all the samples. 

Now we discuss the physical origin of these behaviors. 
The scaling $\rho_{xy} \propto \rho^2_{xx}$ itself is often 
observed experimentally \cite{Hall} and is even derived considering extrinsic mechanisms\cite{side}. In  these conventional theories, a simple proportionality relation $\rho_{xy} \propto M$ is derived in terms of the  perturbative expansion in the spin-orbit coupling $\lambda$ and $M$. 
However, this simple relation is violated because the band crossing occurs in the band structure, and $\lambda M$ lifts this degeneracy\cite{nagaosa,Fang}. This degeneracy point is known to act as a monopole for the gauge field representing the Berry phase curvature,
producing its singular distribution. This non-perturbative feature causes the rapid change of $\sigma_{xy}$ including sign reversal  as a function of $M$ because the Fermi energy crosses this monopole
energy as $M$ changes. First-principles calculations confirm this
scenario as shown in the inset of Fig.~\ref{figSXYT} for SRO (undoped case), 
using the orthorhombic structure obtained for the bulk SRO crystal. 
As seen in this inset, the calculations reproduce closely the non-monotonous 
variation, as well as the sign change of $\sigma_{xy}$ with $M$ (or $T$). 
The first-principles calculations (bottom panel of Fig.~\ref{figSXYT}) show 
how the anomalous Hall conductivity depends on the crystal structure and 
lifetime of the electrons. Results obtained by considering a cubic 
structure are indeed quite different from those obtained in the orthorhombic 
case, even though they qualitatively show a similar non-monotonous behavior. 
It is also shown in the figure how $\sigma_{xy}$ is reduced upon increasing 
the scattering rate in the calculations; the broadening parameters were 
chosen so as to reflect the increase of longitudinal resistivity shown in 
Fig.~\ref{figRT}. The Ca doping of the Sr$_{1-x}$Ca$_{x}$RuO$_3$ films also 
induces slight structural changes, such as a smaller orthorhombicity 
(or more correctly tetragonality, c.f. inset of Fig.~\ref{figRT}). It is 
thus expected that the temperature dependence of $\sigma_{xy}$ of the doped 
films should differ, more or less, from that of the undoped SRO film, 
reflecting the changes in the local lattice-structure of the system and their 
effect on the band structure. Nevertheless, if the additional electron 
scattering effects arising from the Ca doping are taken into account as the 
broadening-induced reduction of the magnitude of $\sigma_{xy}$, the anomalous 
Hall conductivity of the Sr$_{1-x}$Ca$_{x}$RuO$_3$ films shows a good 
scaling to $M$, while changing $T$ and $x$. This indicates that the AHE is 
mainly of intrinsic origin, as described in terms of the Berry phase 
connection.\cite{Fang}

In summary, the anomalous Hall effect was investigated for thin films of 
Sr$_{1-x}$Ca$_{x}$RuO$_3$, in which the ferromagnetic interaction is 
weakened with increasing Ca content $x$. The Hall resistivity of the films 
vary in a similar fashion with the temperature $T$. The obtained anomalous 
Hall conductivity varies non-monotonously and non-trivially with $T$, and 
even changes sign. The results however, show a good scaling solely to the 
$T$- and $x$-dependent magnetization $M$, which can be reproduced by first-principles calculations. The anomalous Hall effect appears, as asserted by 
Fang et al\cite{Fang}, as a hallmark of the presence of magnetic monopoles 
in the momentum space of the crystal.

Z. F. acknowledges the support from the NSF of China (N. 10334090 and 90303022).

\end{document}